\documentclass[a4paper]{spie} 
\usepackage{graphicx}
\usepackage{amsfonts,amssymb,amsbsy,amsmath}

\title{Noiseless amplification of weak coherent fields without external energy} 
\author{Mikko Partanen, Teppo H\"ayrynen, Jani Oksanen, and Jukka Tulkki\skiplinehalf
Department of Biomedical Engineering and Computational Science,\\Aalto University, P.O. Box 12200, 00076 Aalto, Finland}

\begin{document} 

\maketitle 

\begin{abstract}
According to the fundamental laws of quantum optics, noise is necessarily added to the system when one
tries to clone or amplify a quantum state. However, it has recently been shown that the quantum
noise related to the operation of a linear phase-insensitive amplifier can be avoided when
the requirement of a deterministic operation is relaxed. Nondeterministic noiseless linear amplifiers
are therefore realizable. Usually nondeterministic amplifiers rely on using single photon sources.
We have, in contrast, recently proposed an amplification scheme in which no external energy is
added to the signal, but the energy required to amplify the signal originates from the stochastic
fluctuations in the field itself. Applying our amplification scheme, we examine the amplifier gain
and the success rate as well as the properties of the output states after successful
and failed amplification processes. We also optimize the setup to find the maximum success rates
in terms of the reflectivities of the beam splitters used in the setup. In addition, we discuss
the nonidealities related to the operation of our setup and the relation of our setup with the previous setups.
\end{abstract}

\section{Introduction}
Quantum noise is unavoidably added into the signal in any conventional amplification process \cite{Caves1982}.
This follows from the linearity and unitary evolution of quantum mechanics and guarantees against unphysical
situations violating the Heisenberg uncertainty principle. However, it has been shown that the addition of
quantum noise can be circumvented by implementing the amplification process nondeterministically
\cite{Zavatta2011,Ferreyrol2010,Ferreyrol2011,Barbieri2011}. This results in a process in which
conditional quantum operations, such as a sequence of single-photon addition and subtraction, are
applied to the optical field \cite{Marek2010,Hayrynen2009,Hayrynen2011}. Typically, the addition
and subtraction of photons are experimentally implemented by using single-photon light sources,
beam splitters, and photodetectors \cite{Zavatta2011,Ferreyrol2010,Ferreyrol2011,Barbieri2011}.
We have, in contrast, recently proposed an amplification setup in which, the energy required
to amplify the signal does not originate from external energy sources, such as single-photon
sources, but the required energy comes from the stochastic fluctuations in the field
itself \cite{Partanen2012}. In quantum communications and metrology, the noiseless
amplification schemes could become an essential tool to recover information transmitted
over lossy channels or to enhance the discrimination between partially overlapping quantum
states \cite{Zavatta2011,Josse2006}.

In this paper, we analyze our recently proposed amplification scheme, which adds no external energy to
the optical signal. The scheme consists of beam splitters and photodetectors that are routinely used
in experiments and a quantum nondemolition (QND) measurement \cite{Grangier1998,Brune1990,Milburn1984}
that is not as common but nevertheless has several experimental implementations \cite{Munro2005,Nogues1999,Guerlin2007}.
We start by a short summary of the basic principles of noiseless amplification and by a description of
our amplification setup. By optimizing the setup, we find the maximum success rates in terms of the
reflectivities of the beam splitters used in the setup. This is followed by a discussion of the
nonidealities and the realizatization of the QND measurement related to the operation of our setup.
We also investigate the statistics of the output states after successful and failed amplification
processes. Furthermore, we discuss the relation of our setup with the previous setups.

\section{Amplification scheme} 
\begin{figure}
\centering
\includegraphics[width=11cm]{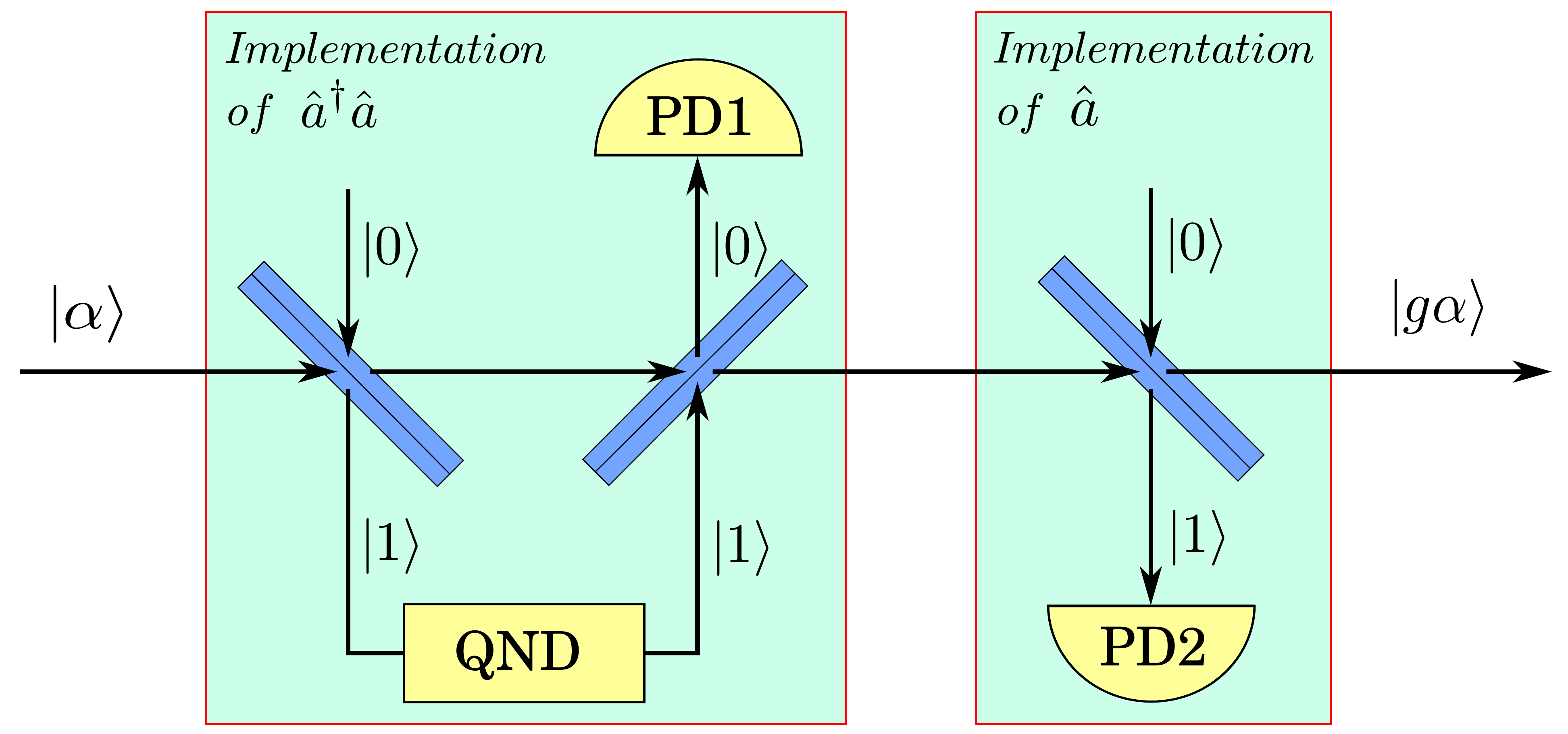}
\caption{\label{fig:scheme} (Color online) A schematic illustration of the noiseless amplification of a weak coherent field. First,
a single photon is subtracted from the field, then added back to the field, and finally again subtracted from the field.
This sequence is also described by the operator $\hat a\hat a^\dag\hat a$.}
\end{figure}

The proposed amplification scheme, illustrated in Fig.~\ref{fig:scheme}, utilizes the energy fluctuations
of the initial field to replace the single-photon source that would otherwise be needed as in the scheme
suggested by Zavatta \emph{et al.} \cite{Zavatta2011}. In our scheme, a similar action is obtained by
a configuration where the successful subtraction of a single photon from the initial field by a beam
splitter is verified by a QND measurement, which is followed by adding the photon back to the field
by the second beam splitter if no photons are detected at photodetector PD1. Finally, another photon
is subtracted from the field at the third beam splitter, if photodetector PD2 detects a photon.
The final output state resulting from these events is an amplified coherent state with high fidelity,
but this output state only occurs when the QND, PD1, and PD2 detectors detect 1, 0, and 1 photons, respectively.

The action of an ideal noiseless amplifier for coherent states can be described as $|\alpha\rangle\rightarrow|g\alpha\rangle$,
where $|\alpha\rangle$ is the initial coherent field, $|g\alpha\rangle$ is the amplified field, and $g$ is the
gain of amplification. This operation cannot be implemented by conventional amplifiers, but it can be
approximated nondeterministically. It has been shown that the operator $\hat G=\hat a\hat a^\dag$, where
$\hat a$ and $\hat a^\dag$ are the annihilation and creation operators of the field, approximates the
action of amplification for weak coherent fields with nominal gain $g=2$ \cite{Fiurasek2009}. 
The scheme suggested by Zavatta \emph{et al.} \cite{Zavatta2011} is based on this.
The same outcome is also obtained by an operator $\hat G^{\,\prime}=\hat a\hat a^\dag\hat a$
implemented by the setup used in this paper because the coherent input field $|\alpha\rangle$ is
an eigenstate of $\hat a$.

The main difference between the presented scheme and the scheme succested by Zavatta \emph{et al.}
\cite{Zavatta2011} is the QND measurement. In general, a QND measurement is a detection process, in
which the detected photon is not destructed as in the case of conventional photodetectors. Thus,
the process should be repeatable. As stated by the laws of quantum mechanics, the measured state
is projected into an eigenstate corresponding to the measurement result. QND measurements can be used
as very sensitive probes of small perturbations acting on a system. In addition, QND measurements
are also used for unlimited distribution of optical signals \cite{Haroche2006}. Various schemes for
QND measurements with different physical systems have been experimentally realized
\cite{Munro2005,Nogues1999,Guerlin2007,Grangier1998,Brune1990,Milburn1984}. For example,
Guerlin \emph{et al.} have observed a step-by-step decay of a cavity field by non-destructively
measuring the photon number of a field stored in a cavity \cite{Guerlin2007}. More recently,
the same group completed another experiment of field damping and measurement of Fock
state lifetimes by QND photon counting in a cavity \cite{Brune2008}. A closely
related work has also been performed on a superconducting quantum circuit \cite{Wang2008}.

\subsection{Output field of the amplifier}
The output fields of our setup have been calculated using the standard Wigner function formalism.
The Wigner function of the initial coherent field $|\alpha\rangle$ is \cite{Schleich2001}
\begin{equation}
W_{\mathrm{coh}}(x,p)=\frac{1}{\pi\hbar}\exp\!\Big[-(\kappa x-\sqrt{2}\mathrm{Re}\alpha)^2-\left(\frac{p}{\hbar\kappa}-\sqrt{2}\mathrm{Im}\alpha\right)^2\Big],
\label{eq:W_coherent}
\end{equation}
where $x$ and $p$ are position and momentum quadratures of the field, $\alpha$ is a complex variable defining the
coherent field amplitude, $\kappa$ is the spring constant of the field oscillator, and $\hbar$ is
the reduced Planck constant. When plotting the Wigner functions, it is conventional to set $\hbar=\kappa=1$
\cite{Schleich2001}.

The entangled Wigner function $W_\mathrm{BS}$ emerging as a result from fields $W_\mathrm{field 1}$
and $W_\mathrm{field 2}$ interfering on a beam splitter is given by \cite{Leonhardt1997,Leonhardt2003}
\begin{equation}
W_\mathrm{BS}(x_1,p_1,x_2,p_2) = W_\mathrm{field 1}(tx_1+rx_2,tp_1+rp_2) W_\mathrm{field 2}(tx_2-rx_1,tp_2-rp_1),
\label{eq:bs}
\end{equation}
where $x_1$, $p_1$, $x_2$, and $p_2$ are the position and momentum quadratures of the transmitted and reflected fields
and the beam splitter reflection and transmission coefficients $r$ and $t$ obey the relation $r^2+t^2=1$.
In our notation $W_\mathrm{field 1}$ is the field incident to the beam splitter from the left and transmitted
field quadratures refer to the field emerging from the beam splitter to the right in Fig.~\ref{fig:scheme}.

The probability of detecting $n$ photons on the reflected field (the field that emerges from the beam splitter
and travels vertically in Fig.~\ref{fig:scheme}) can be expressed as
\begin{equation}
P(n)  = 2\pi\hbar\int W_\mathrm{BS}(x_1,p_1,x_2,p_2) W_n(x_2,p_2)\,dx_1\,dp_1\,dx_2\,dp_2,
\label{eq:probability}
\end{equation}
where $W_n$ is the Wigner function of the $n$-photon Fock state $|n\rangle$, to which the reflected field collapses
after the detection, and is expressed as \cite{Schleich2001}
\begin{equation}
W_n(x,p) = \frac{(-1)^n}{\pi\hbar}\exp\!\Big[-(\kappa x)^2-\left(\frac{p}{\hbar\kappa}\right)^2\Big] L_n\Big[2(\kappa x)^2+2\left(\frac{p}{\hbar\kappa}\right)^2\Big],
\label{eq:W_fock}
\end{equation}
where $L_n(x)$ denotes a Laguerre polynomial of degree $n$. After detecting $n$ photons on the reflected field,
the transmitted field collapses to
\begin{equation}
W_\mathrm{T}(x_1,p_1) = \frac{2\pi\hbar}{P(n)}\int W_\mathrm{BS}(x_1,p_1,x_2,p_2) W_n(x_2,p_2)\,dx_2\,dp_2.
\label{eq:transmitted}
\end{equation}
The collapsed transmitted field is then used as the input for the second beam splitter. Despite the physical
difference between the QND and PD, their effect on the transmitted field is exactly the same, and to calculate
the final output of the setup Eqs.~(\ref{eq:bs})--(\ref{eq:transmitted}) are applied for the remaining two beam
splitters as described in more detail below. For simplicity, we have made the usual assumption that the
photodetectors PD1 and PD2 are ideal. The same assumption is also made for the QND since measurements made
with QND detectors have been reported to yield single-photon Fock states with good accuracy
\cite{Munro2005,Nogues1999,Guerlin2007,Grangier1998,Brune1990,Milburn1984}.
We will discuss the nonidealities more thoroughly in section \ref{sec:nonidealities}.

The details of the analysis of how the state is propagated through the setup resulting in the conditional
state of interest are as follows. In the first beam splitter, the initial coherent field $|\alpha\rangle$
in Eq.~\eqref{eq:W_coherent} is mixed with a vacuum state $|0\rangle$ [a zero-photon Fock state $n=0$
in Eq.~\eqref{eq:W_fock}] using Eq.~\eqref{eq:bs}. Then, one photon is measured by the QND detector.
The probability for this and the transmitted field are given by Eqs.~\eqref{eq:probability} and
\eqref{eq:transmitted} with $n=1$. In the second beam splitter, the transmitted field is mixed with
a single-photon Fock state using Eq.~\eqref{eq:bs} since a photon coming from the QND detector is
added to the field. No photons are measured by photodetector PD1. The probability for this and the
transmitted field are given by Eqs.~\eqref{eq:probability} and \eqref{eq:transmitted} with $n=0$.
In the third beam splitter, a photon is subtracted from the field. This is performed by mixing the
field with a vacuum state using Eq.~\eqref{eq:bs} and using Eqs.~\eqref{eq:probability} and
\eqref{eq:transmitted} with $n=1$ for calculating the probability and the transmitted field that
is the final output state of the setup. The total probability for this successfully amplified output
state $P_\mathrm{succ}$ is the product of the mentioned three photon detection probabilities given by
\begin{equation}
P_\mathrm{succ}=(1+|t_1t_2t_3\alpha|^2(3+|t_1t_2t_3\alpha|^2))
|r_1r_2r_3\alpha|^2 e^{|t_1t_2t_3\alpha|^2-|\alpha|^2},
\label{eq:optfunc}
\end{equation}
where $r_i$ and $t_i$, $i=1,2,3$, are reflectivities and transmittivities of the beam splitters in the setup
obeying $r_i^2+t_i^2=1$.

\subsection{Effective gain and fidelity of the amplified state}
In the calculations depending on the parameters of the setup, effective gain values different from the
nominal gain of 2 can be found. The effective gain can be defined as the ratio of the expectation
values of the annihilation operator $\hat a$ for the output and input fields \cite{Zavatta2011}
\begin{equation}
g_\mathrm{eff}=\frac{|\langle\hat a_\mathrm{out}\rangle|}{|\langle\hat a_\mathrm{in}\rangle|},
\label{eq:geff}
\end{equation}
which corresponds to the effective amplification of the electric-field amplitude.
In the Wigner function formalism, the expectation value of the annihilation operator can be calculated using
the operator correspondence relation as \cite{Gardiner2004}
\begin{equation}
 \langle\hat a\rangle = \int\left[\frac{\kappa}{\sqrt{2}}\left(x+\frac{i\hbar}{2}\frac{\partial}{\partial p}\right)
 +\frac{i}{\sqrt{2}\hbar\kappa}\left(p-\frac{i\hbar}{2}\frac{\partial}{\partial x}\right)\right] W(x,p)\,dx\,dp.
 \label{eq:annih}
\end{equation}
The calculations produce the following expression for the effective gain:
\begin{equation}
g_\mathrm{eff}=\frac{t_1 t_2 t_3 (2+4|t_1 t_2 t_3\alpha|^2+|t_1 t_2 t_3\alpha|^4)}
{1+3|t_1 t_2 t_3\alpha|^2+|t_1 t_2 t_3\alpha|^4}.
\label{eq:gefffull}
\end{equation}

It is also useful to quantify how much the output state differs from an ideally amplified coherent state.
A practical measure for this purpose is the fidelity $F$, which is the overlap between
the states calculated using Wigner functions $W_1$ and $W_2$ of the compared fields \cite{Lee2000}
\begin{equation}
F(W_1,W_2)=2\pi\hbar\int W_1(x,p)W_2(x,p)\,dx\,dp.
\label{eq:fidelityw}
\end{equation}
The fidelity obtained for the successfully amplified field with respect to a coherent
field $|g_\mathrm{eff}\alpha\rangle$ is
\begin{equation}
F_\mathrm{eff}=\frac{(1+2g_\mathrm{eff}t_1t_2t_3|\alpha|^2+g_\mathrm{eff}^2t_1^2t_2^2t_3^2|\alpha|^4)
 e^{-(g_\mathrm{eff}^2-t_1t_2t_3)^2|\alpha|^2}}
{1+3|t_1 t_2 t_3\alpha|^2+|t_1 t_2 t_3\alpha|^4}.
\label{eq:fidelityeff}
\end{equation}

\begin{figure}
\centering
\includegraphics[width=11cm]{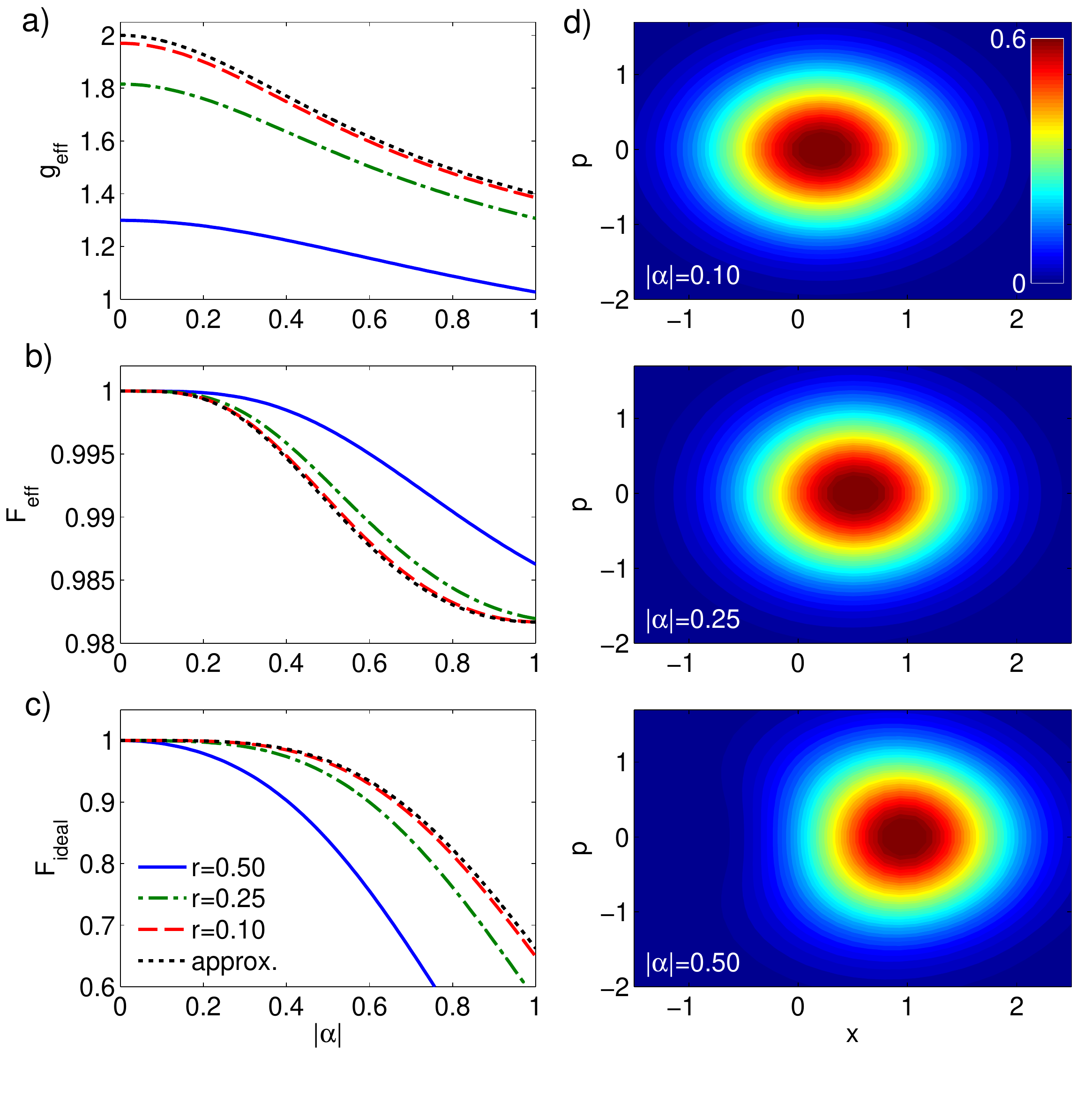}
\caption{\label{fig:succ} (Color online) (a) Effective gain as a function of the input field amplitude for three different
beam splitter reflectivities and for the analytic low reflectivity approximation used by Zavatta \emph{et al.}
\cite{Zavatta2011}. (b) The effective fidelity calculated with respect to a coherent field $|g_\mathrm{eff}\alpha\rangle$.
(c) The fidelity calculated with respect to an ideally amplified field $|2\alpha\rangle$ for comparison
with the results obtained by Zavatta \emph{et al.} \cite{Zavatta2011}. (d) The contour plots of the Wigner functions
for three amplified coherent fields with different input amplitudes when the beam splitter reflectivity is $r=0.4$.}
\end{figure}

The effective gain, fidelity, and Wigner functions of successfully amplified fields are presented in
Fig.~\ref{fig:succ} as a function of the input field amplitude. In Fig.~\ref{fig:succ}(a) the effective
gain is very close to the nominal gain value $g=2$ for small values of $|\alpha|$ and $r$. As the input field
amplitude or the beam splitter reflectivity increases, the gain decreases. Increasing the input field
amplitude results in the reduction of the effective fidelity $F_\mathrm{eff}$ as shown in Fig.~\ref{fig:succ}(b),
where the fidelity is calculated with respect to a coherent field $|g_\mathrm{eff}\alpha\rangle$.
However, the reduction of fidelity can be partly compensated by increasing the beam splitter reflectivity.
Figure \ref{fig:succ}(c) shows the fidelity calculated with respect to an ideal maximally amplified coherent
field $|2\alpha\rangle$ for comparison with the results obtained by Zavatta \emph{et al.} \cite{Zavatta2011}
for the setup, including a specific single-photon source. The values for this ideal fidelity $F_\mathrm{ideal}$
decrease faster than the effective fidelities $F_\mathrm{eff}$ in Fig. \ref{fig:succ}(b) due to the
reduction in $g_\mathrm{eff}$ for stronger input fields. Thus $F_\mathrm{eff}$ is a better measure for the
quality of the resulting output field. One can also see that $F_\mathrm{ideal}$ decreases when the beam splitter
reflectivity increases while the opposite is true for $F_\mathrm{eff}$. This is also due to the reduction
in the effective gain. The contour plots in Fig.~\ref{fig:succ}(d) demonstrate how the Wigner function
deforms in the amplification. For small initial field amplitudes, the output field is very close
to a pure coherent field, but it increasingly deviates from a coherent state when the initial
field amplitude increases.

\subsection{Optimizing the scheme}
Next we investigate how to optimize the probability of successful amplification while maintaining a given
effective gain. The optimization problem for maximizing the probability of successful amplification
[Eq.~\eqref{eq:optfunc}] with a constraint requiring the effective gain [Eq.~\eqref{eq:gefffull}]
exceeding a threshold value $g_\mathrm{eff,0}$ can be presented as
\begin{equation}
\max_{g_\mathrm{eff}\ge g_\mathrm{eff,0}} P_\mathrm{succ}(|\alpha|,r_1,r_2,r_3).
\label{eq:optproblem}
\end{equation}
Here, $|\alpha|$ is the input field amplitude and the beam splitter reflectivities are $r_1$, $r_2$, and $r_3$.
The four-dimensional nonlinear optimization problem in Eq.~\eqref{eq:optproblem} was solved using a barrier function method
\cite{Bazaraa2006}. For a certain $g_\mathrm{eff,0}$, one finds a single maximum $P_\mathrm{opt}$ with an optimal
input field amplitude $|\alpha|_\mathrm{opt}$ and beam splitter reflectivities $r_\mathrm{1,opt}$, $r_\mathrm{2,opt}$,
and $r_\mathrm{3,opt}$. The optimization problem was solved multiple times changing the minimum effective gain
parameter $g_\mathrm{eff,0}$. It was found that at the optimum all the beam splitter reflectivities have the
same value $r_\mathrm{i,opt}=r_\mathrm{opt}$, $i=1,2,3$.

\begin{figure}
\centering
\includegraphics[width=11cm]{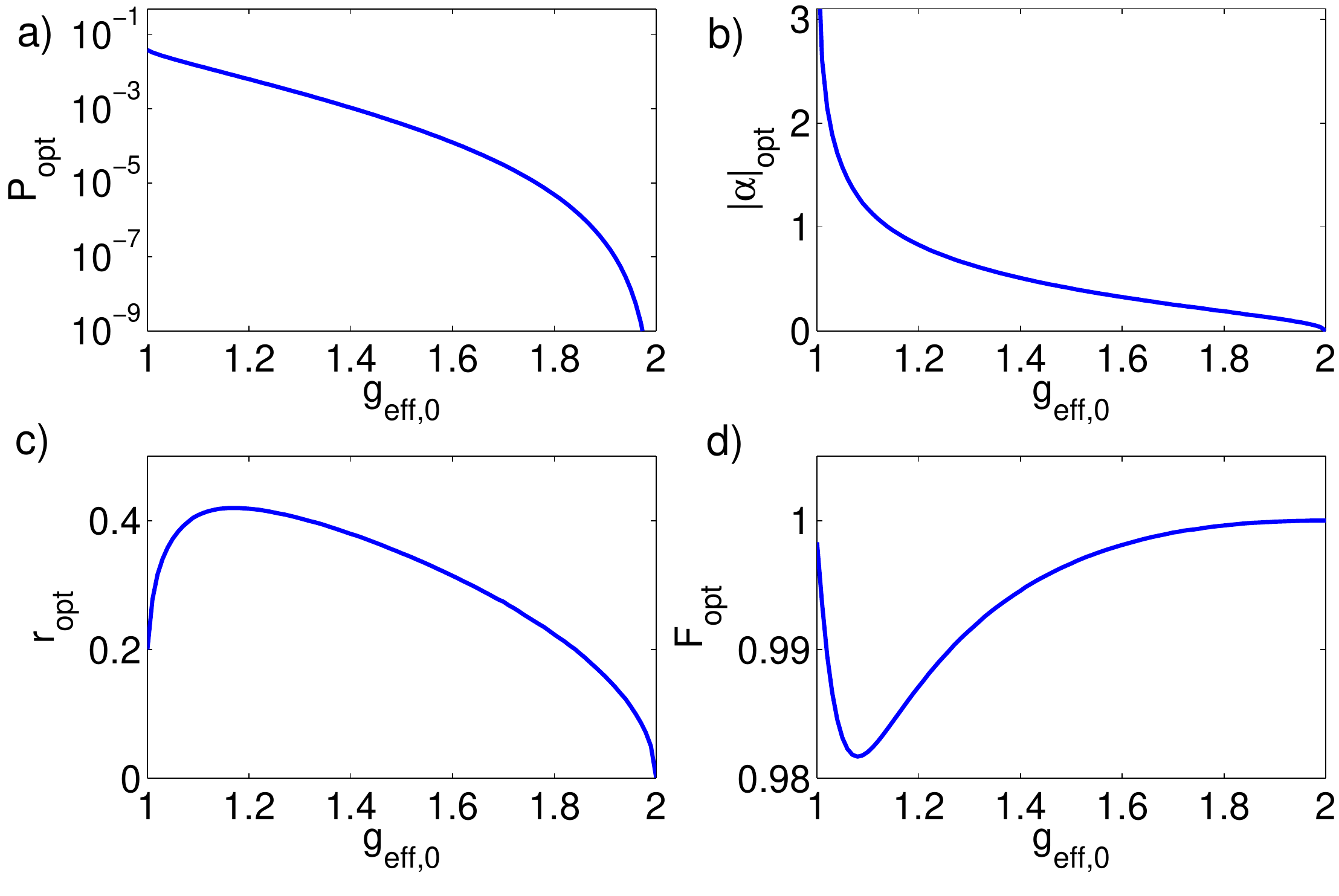}
\caption{\label{fig:opti}(Color online) Probability of successful amplification was maximized in an effective gain constrained
optimization problem. The optimal (a) probability of successful amplification $P_\mathrm{opt}$, (b) the input field
amplitude $|\alpha|_\mathrm{opt}$, (c) the beam splitter reflectivity $r_\mathrm{opt}$, and (d) the output field
fidelity $F_\mathrm{opt}$ are plotted as a function of the required minimum effective gain
parameter $g_\mathrm{eff,0}$.}
\end{figure}

Figure \ref{fig:opti} shows how the optimized
probability of successful amplification $P_\mathrm{opt}$, the input field amplitude $|\alpha|_\mathrm{opt}$,
the beam splitter reflectivity $r_\mathrm{opt}$, and the fidelity of the successfully amplified state
$F_\mathrm{opt}$ evolve as a function of the minimum effective gain parameter $g_\mathrm{eff,0}$.
The probability of successful amplification in Fig.~\ref{fig:opti}(a) decreases exponentially
when the effective gain increases. For instance, if one wants to have an effective gain of 1.4,
the maximum success probability of $10^{-3}$ is achievable with $|\alpha|_\mathrm{opt}=0.51$ and
$r_\mathrm{opt}=0.38$. For comparison, Ferreyrol \emph{et al.} \cite{Ferreyrol2010,Ferreyrol2011} reported
success rates of order $10^{-2}$ for a conventional scheme based on quantum scissors \cite{Ralph2008}.
However, their scheme required a single photon source whose effect is not included in the reported
success rates. Thus the obtained values can not be directly compared.
In Fig.~\ref{fig:opti}(b) one sees that for useful values of $g_\mathrm{eff,0}$ the optimal input field
amplitude is limited to $|\alpha|_\mathrm{opt}<1$. The optimal beam splitter reflectivity in Fig.~\ref{fig:opti}(c)
has a maximum $r_\mathrm{opt}=0.42$ at the effective gain $g_\mathrm{eff,0}=1.18$ and it approaches zero when the
required effective gain approaches 2. The fidelity curve in Fig.~\ref{fig:opti}(d) has a minimum $F_\mathrm{opt}=0.982$
at $g_\mathrm{eff,0}=1.08$. In the nominal gain limit, the fidelity approaches unity.

\newpage
\subsection{\label{sec:nonidealities}QND measurement and other nonidealities}
In the calculations of the probability of successful amplification in equation \eqref{eq:optfunc},
the photodetectors and the QND measurement were assumed to be perfect. In real measurements, this is not
the case, but the obtained success probability must be multiplied with the success probabilities of
individual photodetectors and the QND measurement. Absorptive single photon detection is possible with
efficiencies up to $\sim$90\% for visible spectrum and $\sim$95\% for near-infrared waves
\cite{Hadfield2009,Miller2003,Waks2003}. In addition, Munro \emph{et al.} reported efficiency of $\sim$99\%
for the quantum nondemolition measurement \cite{Munro2005}. Thus, the magnitude of the real success
probabilities is somewhat smaller but still of the same order as the ideal probabilities calculated in this work.

In addition to the success probabilities, one can consider the fidelity of the output states
produced by the setup. In the calculations, it was assumed that the initial state is a perfect
coherent state and in photon addition and subtraction, the detected beam is projected into a
perfect Fock state. In addition, the beam splitters are assumed to be perfect. However, certainly
there are experimental inefficiencies that should be taken into account when calculating the
fidelities of the output states. Conventional photodetectors do not have any effect on the output
state fidelities, but their efficiencies only affect the probabilities of detecting the output
states correctly. In contrast, the QND detector can also affect the output states since the
measured photon is later added back to the field. QND measurements have been reported to project
the field onto Fock states with high fidelity \cite{Brune2008}, but to the authors' knowledge,
no exact values for the fidelities of the produced Fock states have been reported.

\begin{figure}
\centering
\includegraphics[width=11cm]{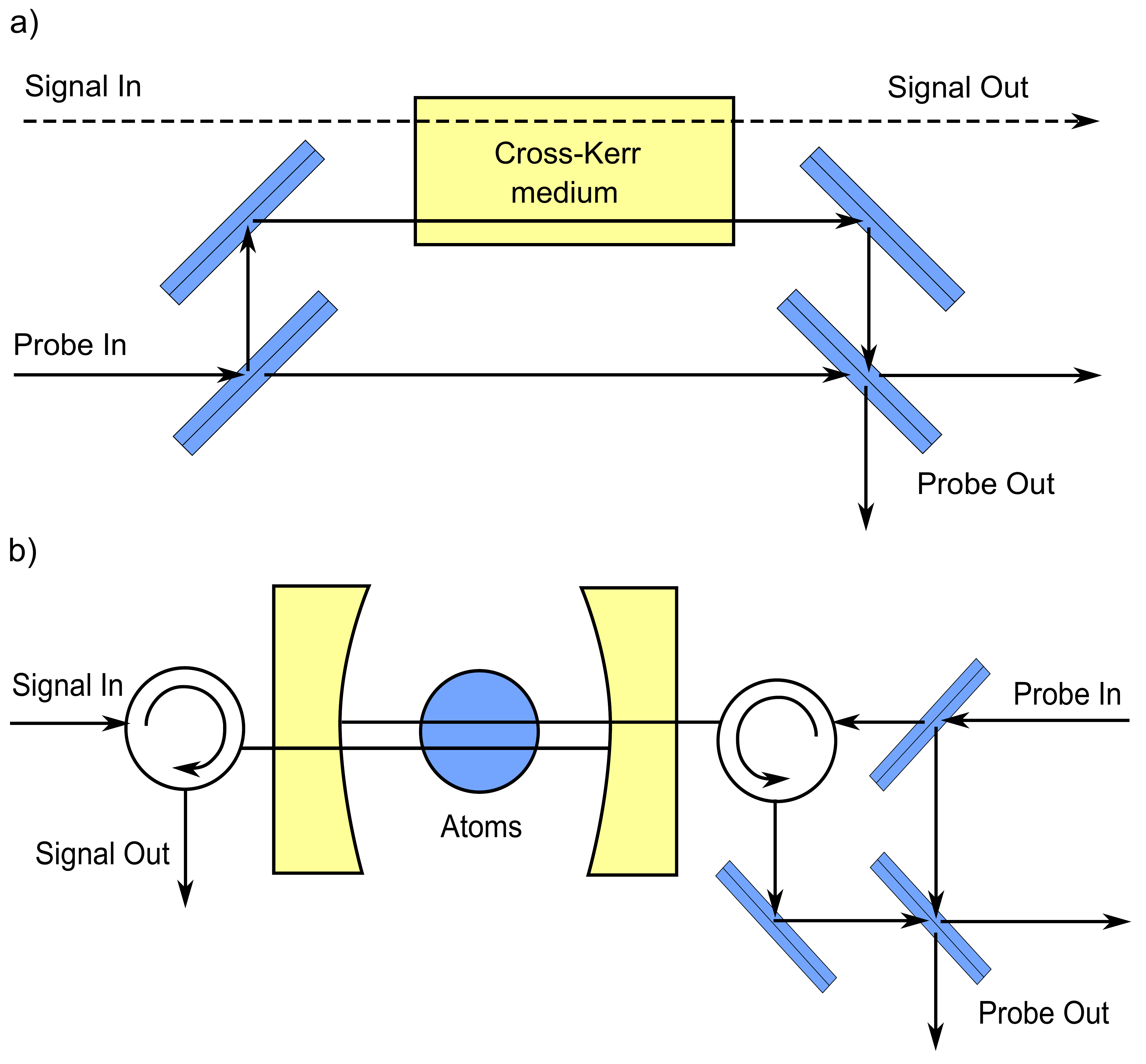}
\caption{\label{fig:qnd}(Color online) Schemes for the implementation of the QND measurement. (a) Measurement
scheme based on cross-Kerr effect \cite{Munro2005}. The intensity of the signal beam alters the refractive
index experienced by the probe beam. This results in a phase shift on the probe beam, based on which the photon
number of the signal beam can be determined. (b) Measurement scheme based on using atoms in an optical cavity
\cite{Nogues1999}. The probe beam experiences a phase shift depending on the photon number of the signal beam
on the cavity. Rotating arrows indicate optical circulators \cite{Grangier1998}.}
\end{figure}

There are at least two suitable implementations for the QND measurement in our amplification scheme:
the scheme based on cross-Kerr effect \cite{Munro2005} and the scheme based on using atoms in an
optical cavity \cite{Nogues1999}. These schemes are presented in Fig.~\ref{fig:qnd}. In the
cross-Kerr nonlinearity based scheme in Fig.~\ref{fig:qnd}(a), the refractive index of the cross-Kerr
medium is altered by the intensity of the signal beam. The probe beam experiences a phase shift
that can be detected. The phase shift is proportional to the photon number of the signal beam,
and thus the photon number of the signal beam can be determined. Also, in the atom and cavity based scheme in
Fig.~\ref{fig:qnd}(b), the phase shift experienced by the probe beam depends on the photon number of
the signal beam, based on which the photon number of the signal beam can be determined.

\newpage
\subsection{Failed amplification}
So far, we have only discussed the case of successful amplification. For completeness, we next analyze the other
possible output states. If the initial field is weak and the beam splitter reflectivities are $<\hspace{-2.5pt}0.5$, the probability
that any of the photodetectors detects more than one photon is typically $<\hspace{-4pt}10^{-2}$. This probability is not completely
negligible but since it is still small and the occurrence of these processes can be detected, we can focus on the
processes where only at most one photon is detected at a time. These single photon processes and the corresponding
eight possible output states are described in Table \ref{tbl:states}. The output states can be experimentally
identified by photon detection measurement outcomes. The first state is the successfully amplified field
and the last row shows the probability that more than one photon is detected by some of the photodetectors.

The fidelities in Table \ref{tbl:states} clearly show that states from 5 to 8 are exactly coherent.
This can be understood by considering what happens if the first photon subtraction fails. In this case, the
output from the first beam splitter can be shown to be $|t\alpha\rangle$, which is a perfectly coherent field.
This further means that at beam splitters 2 and 3, only single-photon subtraction or no photon subtraction
can take place. Both operations result in coherent fields, albeit with reduced amplitude. This is because
the photon subtractions only decrease the amplitude and keep the state coherent since coherent states
are eigenstates of the annihilation operator \cite{Hayrynen2009,Kim2008a}. The states 3 and 4 are not
exactly coherent since, in these cases, the input field arriving to the second beam splitter from the
QND device is a single-photon Fock state producing superposition states at the output.

\begin{table}
\caption{\label{tbl:states}Photon detection measurement outcomes, amplitude expectation values $|\langle\hat a\rangle|$,
degradations of fidelities $1-F_\mathrm{eff}$, and probabilities $P$ for possible single-photon process output states
of the amplification setup with $g_\mathrm{eff}=1.4$. The initial field is a coherent field with $|\alpha|=0.5$ and the reflectivity of
the beam splitters is $r=0.4$. Successful amplification corresponds to the first state.
}
\centering
\renewcommand{\arraystretch}{1.5}
\begin{tabular}{ccccccc}
\hline\hline
        & \multicolumn{3}{c}{Measurements} & & & \\
\cline{2-4}
  State & QND & PD1 & PD2 & $|\langle\hat a\rangle|$ & $1-F_\mathrm{eff}$ & $P$\\
\hline
 1 & 1 & 0 & 1 & 0.686 & $4.84\times 10^{-3}$  & $1.36\times 10^{-3}$\\
 2 & 1 & 0 & 0 & 0.720 & 0.362                & $5.58\times 10^{-3}$\\
 3 & 1 & 1 & 1 & 0.292 & $3.79\times 10^{-5}$  & $5.27\times 10^{-4}$\\
 4 & 1 & 1 & 0 & 0.310 & $1.60\times 10^{-5}$  & $2.88\times 10^{-2}$\\
 5 & 0 & 1 & 1 & 0.385 & 0                    & $8.57\times 10^{-4}$\\
 6 & 0 & 1 & 0 & 0.385 & 0                    & $3.03\times 10^{-2}$\\
 7 & 0 & 0 & 1 & 0.385 & 0                    & $2.55\times 10^{-2}$\\
 8 & 0 & 0 & 0 & 0.385 & 0                    & 0.903\\
 other & & & & & & $3.84\times 10^{-3}$\\
\hline\hline
\end{tabular}
\end{table}

\begin{figure}
\centering
\includegraphics[width=11cm]{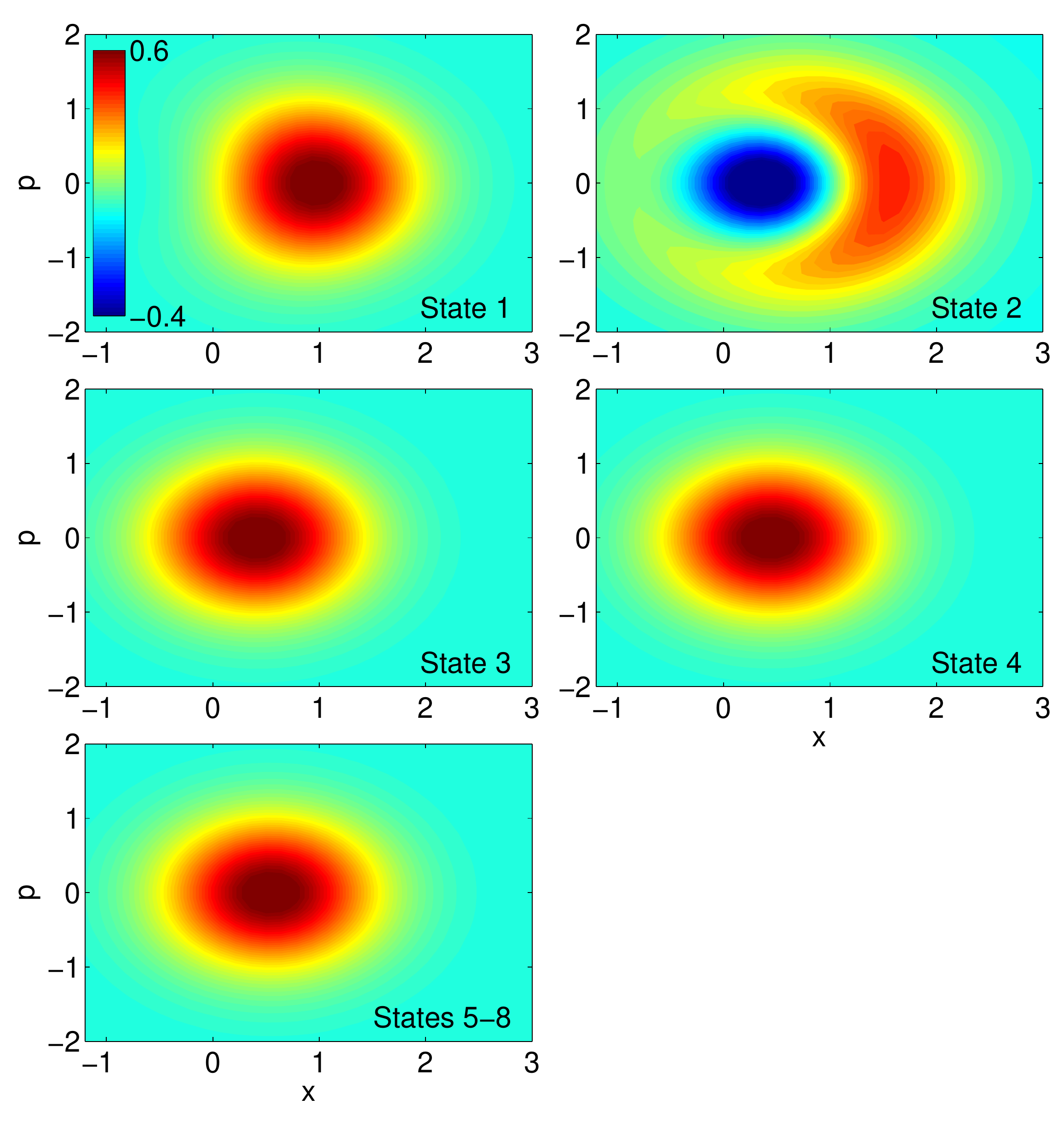}
\caption{\label{fig:states} (Color online) Contour plots of Wigner functions for possible output states when the initial field is
a coherent field with $|\alpha|=0.5$ and the beam splitter reflectivity is $r=0.4$.}
\end{figure}

The contour plots of the different output states are presented in Fig.~\ref{fig:states}. It can be seen that
the only states that notably deviate from the initial coherent field with $|\alpha|=0.5$ are the states 1 and 2 which are
the successfully amplified field and the field after a failure in the last photon subtraction. For other states, one
can hardly find any visible differences.

The amplitude expectation values in Table \ref{tbl:states} showed that the amplitude of the coherent output
states 5--8 $|\langle\hat a\rangle|=0.385$ is clearly smaller than the amplitude of the initial field $|\alpha|=0.5$.
This follows from the relatively large reflectivity of $r=0.4$. For a smaller reflectivity of $r=0.1$, the amplitude
expectation value for the exactly coherent output states is $|\langle\hat a\rangle|=0.493$, which is much closer
to the initial field amplitude. Since this output is also the most probable output and, in the case of small
reflectivities, it is a nearly unchanged coherent state one could also try to repeat the amplification
process in order to increase the probability of successful amplification. However, experimental realization
of the repeated amplification setup would be challenging.

In principle, both the conventional noiseless amplification setups using single photon sources and the proposed
setup relying purely on field fluctuations are based on the same basic concept. They use a device that constructs
a complex superposition output state from the input and collapses the output state into the amplified target state
as a result of certain measurements in the detectors of the device. This unavoidably leads to stochastic operation
where the amplified state can be considered as a fluctuation in the output field, onto which the output state
collapses. However, the proposed setup has also some subtle differences compared with the previously demonstrated
noiseless amplification setups. Most importantly, the amplification in the proposed setup does not add any energy
to the input signal and therefore nicely demonstrates that energy fluctuations in the original signal can be
used as a stochastic energy source for the amplification.

\section{Conclusions}

In conclusion, we have studied noiseless amplification of coherent signals in a setup where all the energy added
to the amplified signal originates from the fluctuations in the quantum field in a purely stochastic manner, i.e.
the field is amplified even when no additional energy is added to the field from external sources in contrast to
the previously reported noiseless amplifiers. We have discussed nonidealities in the setup and the effects
and realizations of the QND detector needed in the setup. We have also shown that the probability of successful
amplification can be maximized by finding optimal values for the beam splitter reflectivities depending on the
desired effective gain. Our results show that the purely stochastic amplification scheme can amplify weak coherent
fields with very good fidelities much like the conventional stochastic amplification setups relying on single
photon sources.

\bibliography{bibliography}
\bibliographystyle{spiebib}

\end{document}